\documentclass[useAMS,usenatbib,referee]{mn2e}

\usepackage{graphicx}
\usepackage{epsf}
\usepackage{fleqn}

\title{Lens candidates in the Capodimonte Deep
Field in vicinity of the CSL1 object}
\author[Sazhin Mikhail]
 { Sazhin M.V.$^1$, Khovanskaya O.S.$^{1}$, Capaccioli M.$^{2,3}$,  Longo G.$^{3,4}$,
\newauthor
Alcal\'a J.M. $^{2}$, Silvotti R.$^{2}$, Pavlov M.V.$^2$\\
 1 - Sternberg Astronomical Institute, Moscow State University,University pr. 13, Moscow, RUSSIA\\
 2 - INAF - Osservatorio Astronomico di Capodimonte, via Moiariello 16, 80131, Napoli ITALY\\
 3 - Department of Physical Sciences, University of Napoli Federico II, via Cinthia 9, 80126 Napoli, ITALY\\
 4 - INFN - Napoli Unit, Dept. of Physical Sciences, via Cinthia 9, 80126, Napoli, ITALY}
\date{Accepted ;
      Received ;
      in original form }
\pagerange{\pageref{firstpage}--\pageref{lastpage}} \pubyear{1994}

\begin{document}

\maketitle

\label{firstpage}

\begin{abstract}
CSL1 is a peculiar object discovered in the OACDF. Photometric and
spectroscopic investigation strongly suggest that it may be the first
case of gravitational lensing by cosmic string. In this paper we
derive and discuss a statistical excess of a gravitational lens
candidates present in OACDF region surrounding CSL1. This excess
cannot be explained on the basis of conventional gravitational lens
statistic alone, but is compatible with the proposed cosmic string
scenario.
\end{abstract}


\section{Introduction}

Cosmic strings in cosmology were thoroughly discussed over the past
decades (cf. \citet{zel}, \citet{vil}). For instance, several authors 
proposed to explain the quick variability of QSO by the gravitational 
lensing by a cosmic string: \cite{sazhin}, \cite{schild}. 

Accordingly to \citet{all}
their expected number should be large enough to make them observable
but, so far, most attempts to detect their signatures seem to have
failed. In a previous paper \citet{csl1} (hereafter Paper I) we
discussed the strange properties of the gravitational lens CSL1
which, by photometric and spectroscopic investigation was
shown to consist of two identical images of a giant elliptical
galaxy at $z = 0.46$. The most relevant feature of these images is
that their isophotes appear to be undistorted, thus leaving only two
possible explanation: the first one is that we are dealing with the
chance alignment of two galaxies and the second one is that CSL1 is
produced by a gravitational lens which does not distort extended
images. 
It has in fact to be stressed that usual gravitational
lenses, i.e. created by a bound clump of matter, produce
inhomogeneous gravitational fields which always distort background
extended images (cf. \citet{gl}, \citet{kee}). The only possible
interpretation of CSL1 in the framework of the gravitational lensing
theory is therefore that of lensing by a cosmic string. In the same
paper we also pointed out that the safest way to disentangle between
these two possible scenarios would be to obtain milliarcsecond
resolution images of CSL1. While waiting for this type of data to
become available, we explore here the possibilities offered by
another method which has been discussed in the past by
\citet{string1}, \citet{string2}, \citet{string3}, \citet{string4},
and recently in \citet{string5}.

The aligment of the background object (a galaxy) inside the
deficit angle of the string is a stochastic process determined by
the area of the lensing strip and by the surface density
distribution of the extragalactic objects which are laying behind
the string. The larger field of search we choose, the larger the
number of lensed objects that we should find. All lensed objects will
fall inside a narrow strip defined by the deficit angle
computed along the string pattern. With an effective metaphor, we
can say that this method calls for the creation of a "new milky
way" of galaxies. consisting of double and triple images, all located
along the string pattern.

The exact number of expected lensed object will be determined by the
deficit angle and geometry of the string as well as by the density
of the extragalactic objects. In what follows we shall consider
only galaxies since the QSO contribution may be shown to be
negligible of optical wavelenghts.

In the case of a straight string one can easily estimate the
expected number of lensed galaxies as

\begin{equation}
\left< N \right> = n_g 2 \delta l
\label{eq1}
\end{equation}

\noindent here $n_g$ is density of galaxies per unit solid angle,
$\delta$ is deficit angle of the string, and $l$ is length of the
string in the chosen field. Both $\delta$ and $l$ are expressed as
angular measures.

A more sophisticated case emerges if the string is assumed to be
curved \citet{string5}. A simple estimation can be derived as
follows. The lenght of a curved string is larger in comparison to a
straight one. Therefore strip of the string will cover a larger area
on the sky in the same patch. The lenght can be written as
\citet{string5}:
\[ l
= R \left(\frac{R}{l_{c}}\right)^a
\]

\noindent
Here $R = |\vec r - \vec r_1|$ is  distance from the point $\vec r$
to the point $\vec r_1$. $l_c$ is a correlation interval.
The parameter $a$ varies inside 0 (straight
string) and 1 (in the case of random walk of the string); the last
value corresponding to purely brownian motion ($R \sim \sqrt{l}$).
In the case (a=1), the expected number of lenses can be estimated as
follows

\[
\left< N \right> =  2 \frac{\delta}{l_{c}} n_g\Omega
\]

\noindent
where the product of the angular area $\Omega$ of the patch, by the
surface density of galaxies $n_g$ gives the multiplication factor
$N_g$ which is number of galaxies in the patch.

In Paper I and in the present work we take the R  OACDF frame as
reference one. We therefore need to estimate the expected number of
galaxies in the R band down to the OACDF limiting magnitude ($\sim$
24 $m_R$).

Deep counts in the R are practicully absent in the literature. A
moderately deep survey was obtained in the R band \citet{couR}. Other
surveys were obtained (cf. \citet{cou}) of different wavelenght and
therefore need to be interpolated in order to adjust to our case.

Another source of information comes from very deep counts such as
those derived by \citet{hubbledf1}, \citet{hubbledf2} from the Hubble
Deep Field. Also in this case however, there are no R band.

By taking into account all the above data it becomes apparent that
galaxy counts have large uncertainty in the interval of magnitudes
$20 < m < 24$ (with a possible multiplication factor of 10). If we
use the \citet{couR} and extrapolate them to $m = 24$ in the R band
we obtain our estimated total number of galaxies in the interval $20
< m_R <24$ are in our (16\arcmin $\times$ 16\arcmin) field of 2200.
It roughly corresponds to number of extended sources in analized
patch of the OACDF. We would stress that this estimate is in good
agreement with the number of galaxies actually observed in the OACDF
in the same magnitude range.

Therefore by using the above estimation in the case of straight
string one can expect 9 lenses along the string, and in the case of
a random string one can expect a much larger number: up to 200 lenses
depending of the value of $a$.

In this patch one can expect also lenses produced by galaxies or
conventional lenses. The mean density of lenses and its application
to cosmology was discussed in \citet{fuk92}, \citet{koc93},
\citet{chi99}, \citet{ofe03}.

One can estimate the mean number of conventional lenses as the
product of the optical depth due lensing per the number of galaxies
in the field. These estimation (within the uncertainty interval)
provide us with a mean number of $\la$ 2 conventional lenses in our
patch.

\section{Criteria for candidates identification}
\subsection{Simulations}
We performed extensive simulations of the gravitational effects of
a cosmic string on a realistic background galaxy distribution. The
simulation was performed as follows. The field was chosen to be
$1000 \times 1000$ pixels in size with a pixel size of 0\farcs238 in
order to match the scale of the OACDF. We stress, however, that
the adopted scale has no relevant effects on the results of the
simulation. The simulated field therefore covers a region of ca.
4\arcmin $\times$ 4\arcmin and the simulated number of galaxies
is slightly more than 100 (as expected from \citet{hubbledf1} and
\citet{hubbledf2} and confirmed by investigation of the OACDF).

Each galaxy was modeled assuming a de Vaucouleurs $r^{1/4}$
surface brightness profile with an artificial threshold at 10
effective radii. The galaxies were then distributed in the field
varying randomly both their positions and their total apparent
magnitudes. Then we smoothed the simulated image with the measured
OACDF point spread function, and finally we added gaussian noise
with zero mean and the same variance as in the R band of OACDF.

The string was assumed to be straight and to cross the simulated
field in a nearly vertical direction.

When the angular distance of a galaxy image from the string is
smaller than the deficit angle of the string, a lensing event
occurs and, as it was shown in Paper I, when the string happens to
cross a galactic image, a triple source appears.

By comparing different simulated field we found on average 3 to 5
lensing events per field, which in some cases were triple ones.
This result is consistent with the theoretical expectations from
Eq. \ref{eq1}.

\subsection{The case of OACDF}
Then we visually inspected a $4000 \times 4000$ pixels subsection
of the OACDF field centered on CSL1. Accordingly to the above
estimates we expect to find  at least 7-9 lens candidates (an
higher number is expected for curved string geometry).

The OACDF is a deep field observed in four broad bands (U, B, V, and
R) plus 6 narrow bands (effective wavelengths at 753 nm, 770 nm,
791 nm, 815 nm, 837 nm, 914 nm). The description of the OACDF and
frames one can find in \citet{cap2}.

In order to select gravitational lens candidates we assumed the
criteria listed in \citet{gl}: i) two or more images with small
angular separation and ii) the same flux ratio in different bands. We
therefore searched for objects having angular separation inside
the interval  1\arcsec - 4\farcs5. The low boundary being determined
by the resolution on the field, and the upper one resulting from the
above described simulation.

Finally, we introduced the following estimator
\[
e_{RV} =1 - \frac{I_R}{I_r} \cdot \frac{I_v}{I_V}
\]

\noindent where $I_R$ is flux in R band from the brightest object
and $I_r$ is flux in R band from the weakest object, $I_V$ is flux
in V band from the brightest object and $I_v$ is flux in V band
from the weakest object. In fact, we used 8 estimators $e_{RB}$,
$e_{R753}$, $e_{R770}$ etc in order to exploit the information
contained in all available bands.

In the case of images produced by a gravitational lens each
estimator should be equal to zero. But, due to the presence of
experimental errors, in the real data, it may assume slightly
different values. Therefore, in order to have a reliable estimate
of the possible range of variation, we derived the estimator for
CSL1, which we assume (from spectroscopy) to be a confirmed case
of lens.

The estimator may be written as follows:
\[
-2.5\log(1 - e_{RV}) = m_{1R} - m_{2R} + m_{2V} - m_{1V}
\]

\noindent Here  $m_{1R}$, $m_{2R}$,  $m_{2V}$,  $m_{1V}$ are the
brightest and the weakest component in the $R$ and $V$ band,
respectively. It is very easy to find a rough estimate of the
errors on the estimator. As far as the estimator is the sum of 4
independent random values, the total error of the estimator is
approximately twice the error on individual magnitudes, because of
the fact that squares are additive values. Therefore the expected
error on the estimators is approximately 20\% of the original
value.

The visual inspection of the OACDF produced a final list of 11 very
likely candidates. This number is in good agreement with our 
cosmic string scenario.

Most of them are weak: magnitudes in R band
ranging from 19 to 24 mag (as expected by \citet{string5}). We also
wish to stress that all candidates having the brightest component
brighter than 21 are also extended objects.

In Fig. 1 we give, as an example, the spectral energy
distributions derived from OACDF photometry the N1 candidate.The
shapes of the spectral distribution of all other candidates are
very similar to the one shown in Fig.1 within the experimental
errors.

\begin{figure}
\vspace{-15pt}
\includegraphics[width=84mm]{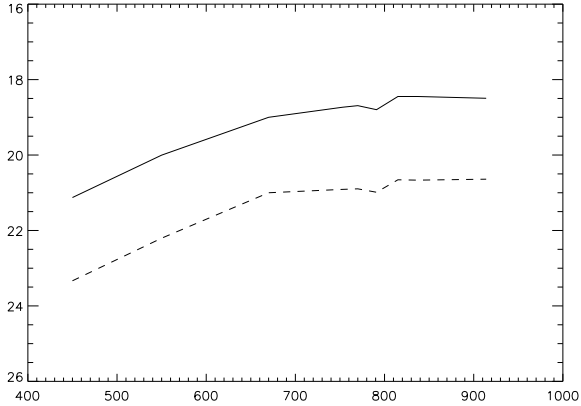}
\vspace{-12pt} \caption{\small Spectral distribution derived from
OACDF multiband photometry. The solid line refers to
the primary object, while the dashed line refer to the secondary
(i.e. fainter) component. The vertical lines give the error bars
for the different spectral bands.} \label{s1}
\end{figure}

It has to be stressed, however, that while in the case of
microlensing, the similarity of photometric colors is the main
criterium, in extragalactic lensing, spectroscopic confirmation is
needed.

\section{The angular separation vs magnitude difference test}

Theory predicts that, in the case of lensing of a point source by
a cosmic string, the angular distance between images should be
roughly equal to the deficit angle, while in the case of lensing
of a background extended object we expect a weak correlation
between the distance of the photometric centroids and the
magnitude difference of the two images. More in detail, in the
case when a small part of the background extended object falls
inside the lensing strip, the weak secondary image forms on the
other side of the string. In this case the distance between the
centroids of the images is the sum of the deficit angle and of the
size of the object itself. Therefore, a correlation between the
ratio of the fluxes and the angular distances between the centers
should arise; the actual shape depending on the exact brightness
distribution within the lensed object.

\begin{figure}
\vspace{-15pt}
\includegraphics[width=84mm]{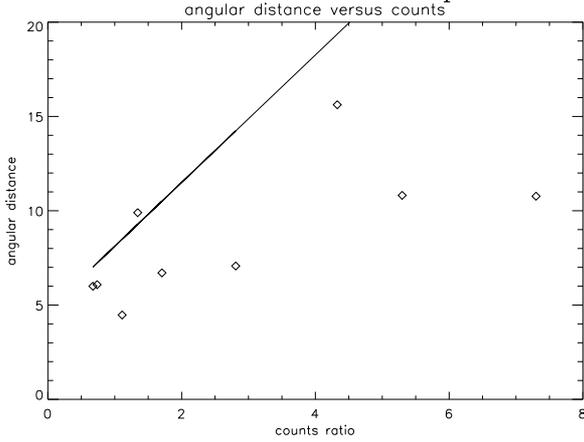}
\vspace{-12pt} \caption{\small Here correlation of angular
distance versus ratio of fluxes of two images is presented.
Diamonds show position of experimental dots, while solid line
present expected theoretical function obtained in simulations}
\label{s2}
\end{figure}

In Fig. \ref{s2} we plot the central peaks distance against the
magnitude difference in the $R$ band for all our lens candidates. The
solid curve gives the trend derived from our simulations, while
diamonds refer to our candidates.  The existence of a weak
correlation for the candidates can be considered as an argument in
favor of the cosmic string model.

\section*{Conclusion}

In Paper I we discussed the strange properties of CSL1: a peculiar
object discovered in the OACDF which spectroscopic investigations
proveed to be the double undistorted image of an elliptical galaxy.
Always in Paper I we showed that CSL1 could be interpreted as the
first case of lensing by a cosmic string.

In the present work, starting from consideration that a cosmic string
is an elongated structure which produces non local effects we
investigated the statistics of lens candidates around the CSL1
position.

The result of extensive simulations led us to expect a relativity
high number of lenses: 7 -9. A prediction which seems to be confirmed
by the investigation of a 16\arcmin $\times$ 16\arcmin area centered
on CSL1. By applying the objective estimator described in the text we
found, in fact, 11 candidates which show the lens properties listed
in \citet{gl}. this figure is much higher than what has to be
expected from normal gravitational lens statistics.

It has to be stressed however that firm conclusions, will be
shown only when spectroscopic confirmations of these candidates will
become avaible.

\section*{Acknowledgments}

M.V.Sazhin acknowledges the Capodimonte Astronomical Observatory for
hospitality and financial support. The work was also supported by 
Russian Fund of Fundamental Investigations No. 04-02-17288.

\label{lastpage}

\end{document}